\documentclass[preprint2,11pt]{aastex}

\lefthead{}
\righthead{}

\begin{document}

\title{No evidence for interstellar planetesimals \\ trapped in the Solar System} 

\author{\textbf{A. Morbidelli$^{(1)}$,  K. Batygin$^{(2)}$, R. Brasser$^{(3)}$, S.N. Raymond$^{(4)}$}} 
\affil{(1) Laboratoire Lagrange, UMR7293, Universit\'e de Nice Sophia-Antipolis,
  CNRS, Observatoire de la C\^ote d'Azur. Boulevard de l'Observatoire,
  06304 Nice Cedex 4, France. (Email: morby@oca.eu) \\
  (2) Division of Geological and Planetary Sciences, California Institute of Technology, Pasadena, CA 91125, USA \\
  (3) Earth Life Science Institute, Tokyo Institute of Technology, Ookayama, Meguro-ku, Tokyo 152-8550, Japan\\
  (4) Laboratoire d'Astrophysique de Bordeaux, CNRS and Universit{\'e} de Bordeaux, All\'ee Geoffroy St. Hilaire, F-33165 Pessac, France
} 

\begin{abstract}
  {In two recent papers published in MNRAS, Namouni and Morais (2018, 2020) claimed evidence for the interstellar origin of some small Solar System bodies, including i) objects in retrograde co-orbital motion with the giant planets, and ii) the highly-inclined Centaurs. Here, we discuss the flaws of those papers that invalidate the authors' conclusions. Numerical simulations backwards in time are not representative of the past evolution of real bodies. Instead, these simulations are only useful as a means to quantify the short dynamical lifetime of the considered bodies and the fast decay of their population. In light of this fast decay, if the observed bodies were the survivors of populations of objects captured from interstellar space in the early Solar System, these populations should have been implausibly large (e.g. about 10 times the current main asteroid belt population for the retrograde coorbital of Jupiter). More likely, the observed objects are just transient members of a population that is maintained in quasi-steady state by a continuous flux of objects from some parent reservoir in the distant Solar System.  We identify in the Halley type comets and the Oort cloud the most likely sources of retrograde coorbitals and highly-inclined Centaurs.
\vskip .5truecm\hbox{}}
\end{abstract}

\section{Introduction}
The passages of the interstellar objects 1I/Oumuamua and 2I/Borrisov through the Solar System on clearly hyperbolic orbits  have stimulated interest {in} extrasolar planetesimals and their similarities and differences with the small bodies of the Solar System. It is therefore not surprising that the claims by Namouni and Morais (2018, 2020) on the existence of populations of extrasolar planetesimals stranded in the Solar System since 4.5 Gy ago {have} attracted some attention in the astronomical community and in the media. Although the willingness of Namouni and M{o}rais to consider unconventional possibilities is admirable, the analys{e}s outlined in the aforementioned papers are not correct. In particular, the logic of the presented arguments suffers from significant drawbacks, and the methods are unsupported by modern knowledge of the behavior of chaotic dynamical systems. Below we summarize the main steps of the Namouni and Morais analysis, then discuss why they are not valid.

\section{A brief summary of the work by Namouni and Morais}

In their 2018 paper, Namouni and Morais consider the object (514107) 2015 BZ509 that is currently on a retrograde {orbit, executing} co-orbital motion with Jupiter. In their 2020 paper they extend their analysis to several other small bodies {in} co-orbital {resonances} with the giant planets and to Centaurs with highly{-}inclined or retrograde orbits. These are the objects that they claim to be of interstellar origin.

Their work can be very simply summarized as follows:
\begin{itemize}
\item They clone the observed objects about a million times (the exact number depends from object to object), with orbital elements sampling the current uncertainty on the nominal orbits of the real objects;
\item They integrate all of the clones backwards in time for 4.5 Gy (the  {approximate} age of the Solar System);
\item They find that the vast majority of the clones don't survive for the whole integration timespan. Most are ejected from the Solar System, or collide with the Sun or the planets. The typical {dynamical} lifetimes are  a few My. Only one clone in a million of (514107) 2015 BZ509 preserves its initial orbital characteristics for 4.5~Gy and only $\sim 1.5\times 10^{-4}$ of the clones of all objects (with the exception of 2008 KV42 and (471325) 2011 KT19)\footnote{The clones of 2008 KV42 and (471325) 2011 KT19 have a higher survival rate and these objects are discussed separately in Sect.~3} are still on orbits bound to the Sun at the end of the integration timespan. Hereafter we refer to these as the ``4.5~Gy surviving trajectories'';
\item They then invoke a very personal view of the {\it Copernican principle} --~according to which we should not be living in a special moment of the history of the Solar System {(Bondi, 1961)}~-- to assert that the real objects must have followed the 4.5~Gy surviving trajectories, even if these are just a strict minority of all possible dynamical outcomes. Otherwise --~the authors {claim}~-- we {must} be living in a special moment when these objects are observable, which would be only a small fraction of the Solar System lifetime;
\item From this, they conclude that the considered objects must be of interstellar origin.  They justify their claim with the following argument. The clones following the 4.5~Gy surviving trajectories are in the end on highly{-}inclined orbits with respect to the plane of the planets, whereas at that time the whole Solar System should have been {shaped like} a disk {(namely, all inclinations should have been small)}.  The considered objects must therefore have originated outside our system, {i.e.} from elsewhere in the Galaxy.
\end{itemize}
We now discuss the problems with this approach.

\section{The fatal flaws}

Although the classical systems studied in Celestial Mechanics conserve energy and momentum, the dream of Laplace to know the past history of a system by integrating its evolution backwards in time with sufficient precision cannot be fulfilled. The reason is that, as demonstrated by Poincar\'e (1899), the systems of Celestial Mechanics are in general non integrable; most of the initial conditions lead to chaotic dynamics (Henon and Heiles, 1964). This is certainly the case for the objects considered by Namouni and Morais (2018, 2020), given their close encounters with the planets and statistically short dynamical lifetimes. A swarm of particles on chaotic trajectories can be described with the tools of statistical mechanics. In particular, their entropy (the exponential of which in this case is related to the phase-space volume occupied by the ensamble of clones) increases with time (Gibbs, 1902). Because of the exponential accumulation of errors in presence of chaotic dynamics, the entropy increases in both forward and backward integrations (Gaspard, 2005). Therefore backward and forward integrations are statistically equivalent{. T}he backward integrations don't reproduce {--}~even in a statistical sense~{--} the real past evolution of the system {because}, if one could follow the {\it real} evolution backward in time, the entropy of the system {would} decrease in agreement with the second law of thermodynamics.

For clarification, consider the following thought experiment. {Suppose t}here is a bottle full of aromatic molecules in a room; the cap is opened and the aromatic molecules diffuse out of the bottle into the room. Simulating this system would not be difficult; the diffusion of the molecules from the bottle into the room corresponds to a net increase in entropy. Now, imagine to come into the room once its air is full of aromatic molecules and to wonder where they come from. The bottle is open in one corner of the room and you wonder whether the aromatic molecules may have come from there. Given that you believe yourself to be the incarnation of Laplace's demon, you measure all the position and velocities of the aromatic molecules and of all other gas molecules in the room and you start simulating their dynamical evolution backwards in time. Given that no measurement is made with infinite precision, the aromatic molecules will never appear to go back all together into the bottle. Therefore, you would conclude --~incorrectly~-- that the bottle was not the source of the perfume.  Now, the objects considered in this discussion are like the aromatic molecules and the Solar System's primordial disk is like the bottle. Then, it should not be surprising that the 4.5~Gy surviving trajectories in the simulations by Namouni and Morais are not found in the disk.

This issue is well known by all dynamical astronomers. Nobody has ever seriously thought to  find the source regions of near-Earth asteroids, Jupiter-family  or long-period comets by integrating their evolution backwards in time. Instead, state of the art models for these populations make an educated guess of the{ir} respective source regions, then simulate {\it forward} in time the evolution of the objects and finally compare the results with the observations validating, in case of success, the initial {\it ansatz} on the source (see Bottke et al., 2002 and Granvik et al., 2018, for the near-Earth asteroids; Levison and Duncan, 1997 and Nesvorny et al., 2017 for the Jupiter family comets; Wiegert and Tremaine 1999 for the long-period comets).

The second serious flaw is in the application of the Copernican principle. The Solar System is an evolving system and therefore there is no reason a priori that the Solar System that we see today is identical to the Solar System in the past. A strict application of the principle as interpreted by Namouni and Morais would simply lead to the statement that the Solar System has always been like it is now. Indeed, such a view is sure to cause intellectual discomfort to anyone who has ever seen a shooting star (a meteor) zoom across the night sky, only to disappear for all time. Moreover, followed to its logical conclusion, this would imply that all objects out of the Solar System{'}s midplane {--} including Pluto and even Mercury {--} must be exogenous! {Furthermore, n}otice that the Copernican principle as stated in Namouni and Morais (2020) is contradicted within the paper itself. In fact, the 4.5~Gy surviving trajectories lead the objects onto radically different orbits than those they occupy today. So, we would be living in a special time to see them on their current orbits.

Leaving behind this philosophical discussion, the truth is that if the dynamical lifetime of a set of observed objects is short, the objects can be the relic of a primordial population only if the latter originally comprised many more bodies. This is for instance the case of the scattered disk in the trans-Neptunian population: the scattered disk objects are unstable and they are believed to be the remnant of a primordial scattered disk which was originally $\sim 100$ times more populated than now (Duncan and Levison, 1997), formed during the period of Neptune's migration (Brasser and Morbidelli, 2013; Nesvorny et al., 2017). With a probability of 1 in a million to remain on its current trajectory  for 4.5~Gy, (514107) 2015 BZ509 should therefore be the remnant of an initial  population of about one million objects of comparable size (about 3-4~km in diameter) on similar orbits {(retrograde, with semi-major axis oscillating around Jupiter's value; Meeus, 2019)}, i.e. about 10 times the current population in the asteroid belt. This is quite implausible given the expected volume density of interstellar planetesimals in the galaxy (Meech et al., 2017; Do and Tucker, 2018).

The third fatal flaw of Namouni and Morais is to neglect a priori the possibility that these strongly unstable objects are transient representatives of a population that is maintained in steady state. {Indeed, to reconcile short dynamical lifetimes of real objects with} the Copernican principle, {a} steady state {scenario} is the {most likely} solution. There are multiple examples of populations of small bodies in the Solar System with individually short lifetimes that are maintained in steady state by {a} flux of new objects from a parent reservoir. The near-Earth asteroid population is a clear example. With a median lifetime of $\sim 10$~My (Gladman et al., 1997) the individual near-Earth asteroids come and go in the blink of an eye compared to the Solar System{'s} age, but are substituted by new objects leaking out of the main asteroid belt (e.g. Morbidelli and Vokrouhlicky, 2003). The coorbital asteroids of the Earth or Venus have a very short residence time ($\sim 25,000$~y) on their characteristic orbits, but their populations as a whole are kept in steady state by the temporary trapping of near-Earth asteroids (Morais and Morbidelli, 2002, 2006). The Jupiter family comets have a combined dynamical/physical lifetimes of $\sim 10^4$~y (Levison and Duncan, 1997), but their population is  kept in steady state by the injection of objects that originate within the scattered disk (Duncan and Levison, 1997). The long-period comets have a lifetime of a few orbital revolutions, but are kept in steady state as a population by the incoming flux of new comets from the Oort cloud (Wiegert and Tremaine, 1999). So, a steady-state scenario should be the default explanation for the existence of short-lived small bodies and one should look for more exotic explanations only if no source capable of maintaining the required steady state is found.

It is not the purpose of this short rebuttal to build steady state models for the objects considered in the Namouni and Morais papers. Nevertheless, a few suggestions can be provided. For (514107) 2015 BZ509, the retrograde coorbital of Jupiter, the population of Halley type comets (themselves coming from the Oort cloud; Fernandez and Gallardo, 1994) could be an obvious source to consider. The inclination of (514107) 2015 BZ509 is well within the range of Halley-type comets and it is well known that during their dynamical evolution short-period comets are often temporary trapped in mean motion resonances with the giant planets. For the high{ly}-inclined or retrograde Centaurs, the obvious source would be the Oort cloud, from which objects come into the inner Solar System with an isotropic distribution of inclinations (Brasser et al., 2012b).  Here a difficulty is that the Centaurs have semimajor axes much smaller than those of the typical comets from the Oort cloud; encounters with the planets can decreases the semi major axes, but if the objects encounter only Uranus and Neptune at high relative velocity, as it is the case for the considered high{ly}-inclined Centaurs, the planetary close encounters may not be very effective. Nesvorny et al. (2019) indeed found a deficit of highly-inclined  Centaurs in their model, although this may be an issue of small number statistics (one of such objects was found in the survey, while the model predicts a 10\% probability of having one detection). Nevertheless the issue needs further analysis. As an alternative explanation, Gomes et al. (2015) and Batygin and Brown (2016) proposed that the high{ly}-inclined Centaurs are one of the signatures of the existence of a putative IXth planet in the distant Solar System.

The objects 2008 KV42 and (471325) 2011 KT19 may require specific consideration. With a median dynamical lifetime of 100-200 My and a 4.5~Gy survival probability of 15\% in the simulations of Namouni and Morais (2020), these objects may in principle be the remnant of an initially large, but not abnormal population, possibly established during the dispersal of the original planetesimal disk in presence of a natal stellar cluster,  which is the scenario invoked for Sedna and the inner Oort cloud (Brasser et al., 2012a). {As an alternative,} Batygin et al. (2019) reproduced their existence (called Niku and Drac in that publication) under the Planet IX hypothesis. Clearly, more investigations are required before we can conclude on the origin of these objects. Nevertheless, their capture from interstellar space is far from obvious. In fact, putative objects trapped from the interstellar space are expected to have orbits typical of the Oort cloud (Levison et al., 2010; Hands and Dehnen, 2020), i.e. radically different from those of 2008 KV42 and (471325) 2011 KT19\footnote{Siraj and Loeb (2019) claimed that interstellar objects trapped in the Solar System by a Jupiter encounter can acquire high{ly}-inclined Centaurs orbits. Nevertheless, the orbits of 2008 KV42 and (471325) 2011 KT19 are more than $10\sigma$ away from the maximum of the probability distribution of the orbital elements of their captured objects. In addition, Hands and Dehnen (2020) reported they could not reproduce Siraj and Loeb's results.}. {In the end,} Namouni and Morais do not present any model reproducing the orbits of the these objects (or any other Centaur) via the capture of interstellar bodies, {meaning that even the basic premise of this scenario remains undemonstrated}. 

\section{Conclusion}
We have discussed in some detail the fatal flaw{s} that invalidate the claims made in Namouni and Morais (2018, 2020) {pertaining to the existence of}  extrasolar planetesimals on bound Solar System orbits. Although it is not strictly impossible that interstellar comets can become temporarily or {even} permanently trapped within the {S}olar {S}ystem, to date, no evidence for their existence has been marshaled. Thus, to study exhotic planetesimals our attention can only turn to 1I/Oumuamua and 2I/Borrisov and to the {other} objects on hyperbolic trajectories that will {undoubtedly} be discovered in large numbers in the future. 

\section{References}

\begin{itemize}
\item[--] Bottke W.~F., Morbidelli A., Jedicke R., Petit J.-M., Levison H.~F., Michel P., Metcalfe T.~S., 2002, Icar, 156, 399
\item[--] Batygin K., Brown M.~E., 2016, ApJL, 833, L3
\item[--] Batygin K., Adams F.~C., Brown M.~E., Becker J.~C., 2019, PhR, 805, 1
  \item[--] {Bondi, H., 1961. {\it Cosmology}, Cambridge University Press.}
  \item[--] Brasser R., Duncan M.~J., Levison H.~F., Schwamb M.~E., Brown M.~E., 2012a, Icar, 217, 1
    \item[--] Brasser R., Schwamb M.~E., Lykawka P.~S., Gomes R.~S., 2012b, MNRAS, 420, 3396
    \item[--] Brasser R., Morbidelli A., 2013, Icar, 225, 40
      \item[--] Do A., Tucker M.~A., Tonry J., 2018, ApJL, 855, L10
  \item[--] Duncan M.~J., Levison H.~F., 1997, Sci, 276, 1670
    \item[--] Fernandez J.~A., Gallardo T., 1994, A\&A, 281, 911
\item[--] Gaspard P., 2005, cdtc.book, 182, 107, cdtc.book
\item[--] Gibbs, J.W. 1902. Elementary Principles in Statistical Mechanics, developed with especial reference to the rational foundation of thermodynamics
\item[--] Gladman B.~J., et al., 1997, Sci, 277, 197
  \item[--] Gomes R.~S., Soares J.~S., Brasser R., 2015, Icar, 258, 37
  \item[--] Granvik M., et al., 2018, Icar, 312, 181
    \item[--] Hands T.~O., Dehnen W., 2020, MNRAS, 493, L59
  \item[--] Henon M., Heiles C., 1964, AJ, 69, 73
  \item[--]  Laplace. P.S. 1814. Essai philosophique sur les probabilit\'es.
  \item[--] Levison H.~F., Duncan M.~J., 1997, Icar, 127, 13
  \item[--] Levison H.~F., Duncan M.~J., Brasser R., Kaufmann D.~E., 2010, Sci, 329, 187
  \item[--] Meech K.~J., et al., 2017, Natur, 552, 378
    \item[--] {Meeus J., 2019, JBAA, 129, 170}
  \item[--] Morais M.~H.~M., Morbidelli A., 2002, Icar, 160, 1
  \item[--] Morais M.~H.~M., Morbidelli A., 2006, Icar, 185, 29
  \item[--] Morbidelli A., Vokrouhlick{\'y} D., 2003, Icar, 163, 120
  \item[--] Namouni F., Morais M.~H.~M., 2018, MNRAS, 477, L117
    \item[--] Namouni F., Morais M.~H.~M., 2020, MNRAS, 494, 2191
    \item[--] Nesvorn{\'y} D., Vokrouhlick{\'y} D., Dones L., Levison H.~F., Kaib N., Morbidelli A., 2017, ApJ, 845, 27
      \item[--] Nesvorn{\'y} D., et al., 2019, AJ, 158, 132
      \item[--] Poincar\'e, H. 1899. Les méthodes nouvelles de la mécanique céleste
        \item[--] Siraj A., Loeb A., 2019, ApJL, 872, L10
      \item[--] Wiegert P., Tremaine S., 1999, Icar, 137, 84

\end{itemize}
\end{document}